\documentclass[twocolumn,showpacs,preprintnumbers,amsmath,amssymb]{revtex4}

\usepackage{graphicx}
\usepackage{dcolumn}
\usepackage{bm}
\begin{document}

\preprint{APS/123-QED}

\title{Phase transition of a one-dimensional Ising model with distance-dependent connections}

\author{YunFeng Chang$^{1}$, Liang Sun$^{1,2}$, Xu Cai$^{1}$}
\affiliation{$^{1}$Complexity Science Center,Institute of Particle
Physics, Huazhong Normal University, Wuhan, 430079, People's
Republic of China\\ $^{2}$Department of Physics and Technology,
University of Bergen, All$\acute{e}$gaten 55, Bergen, 5007,
Norway}

\date{\today}

\begin{abstract}
The critical behavior of Ising model on a one-dimensional network,
which has long-range connections at distances $l>1$ with the
probability $\Theta(l)\sim l^{-m}$, is studied by using Monte
Carlo simulations. Through analyzing the Ising model on networks
with different $m$ values, this paper discusses the impact of the
global correlation, which decays with the increase of $m$, on the
phase transition of the Ising model. Adding the analysis of the
finite-size scaling of the order parameter $[\langle M\rangle]$,
it is observed that in the whole range of $0<m<2$, a
finite-temperature transition exists, and the critical exponents
show consistence with mean-field values, which indicates a
mean-field nature of the phase transition.
\end{abstract}


\maketitle

\section{Introduction}
Since first proposed by Watts and Strogatz \cite{s1} in 1998, the
Watts-Strogatz (WS) small world model has been widely and deeply
studied. In WS model, vertices are placed on a ring with each
vertex having a finite number of $2k$ nearest regular connections
initially. The connections are then rewired with a probability
$p_{r}$ to form long-range connections or short paths. By varying
the single parameter $p_{r}$ from $0$ to $1$, WS model displays
phase transition from a regular network to a small world, which
ends up with a random network at $p_{r}=1$. Considering that the
rewiring of connections may cause isolated vertices, the
addition-type WS model - Newman and Watts (NW) model \cite{s2,s3}
was proposed later, in which long-range connections are added with
a probability $p$ while keeping the initial regular connections
unchanged. NW model and WS model are nearly equivalent, especially
when the network size $N\rightarrow\infty$ \cite{s2,s3}. Marcelo
Kuperman and Guillermo Aberamson generalized WS model in
\cite{s4}. In their model, besides connecting with the two nearest
neighboring vertices, with a probability $p$ each of the vertices
in the model chooses a non-neighboring vertex to construct a
long-range connection between them. The choice of the
non-neighboring vertex is governed by the distance-dependent
probability distribution:

\begin{equation}
\Theta(l)\propto l^{-m}.
\end{equation}

When $m=0$, the NW model is restored. Through varying the value of
$m$, the global correlation of the network is influenced. By
analyzing the topological characteristic of the network, it was
observed that the network also shows a small world nature, and the
network converges to a regular lattice as $m\rightarrow\infty$
\cite{s4,s5,s6,s7}.

To analyze the behavior of a small world network from the point of
view of statistical physics, researchers focus on its long-range
connections or short-paths, which may lead to global coherence. In
networks without any short-path, the transmission of information
has to pass a long distance $\sim O(N)$ ($N$ is the network size),
so global coherence is difficult to reach. With the increase of
the number of short-paths, or long-range connections, collective
behavior will emerge in the system. This global coherence and the
ubiquitous small world phenomenon in real-life make the study of
thermodynamic phase transition on networks with small world
property significantly popular \cite{s8,s9,s10,s11,s12,s13}, in
which Ising model is one of the most fascinating points.

As a comparatively simple but very important model of statistical
physics, Ising model perfectly shows order-disorder transition of
the system. Ising model on one-dimensional regular lattice does
not show phase transition at any finite-temperature and the phase
transition on multi-dimensional regular lattice is not of
mean-field nature. That the Ising model undergoes phase
transitions in the addition-type WS model has been studied in
\cite{s10,s11,s12,s13}. It has also been shown that in the small
world phase the addition-type WS model has a mean-field nature
\cite{s11,s12,s13}. The critical behavior of the Ising model on
one-dimensional network with distance-dependent connections given
by Eq.(1) is expected to reflect the nature of the network at
different values of $m$ indirectly \cite{s12}. But it remains
controversial about the mean-field nature of the phase transition
in the range of $1\leq m<2$ \cite{s6,s7,s12}.

To detect whether a network has a small world behavior, one can
calculate the average shortest-path length and the average
clustering coefficient. Sen and Charkrabarti in \cite{s6} and
Moukarzel and de Menezes in \cite{s7} obtained the contradictory
results by analyzing the topological structure of networks with
connection probability at distance $l$ given by Eq.(1). In Ref.
\cite{s6}, it was found that the average shortest-path length
behaves as $\ln N$ on rings of size $N$ for all $m<2$, hence it
was concluded that small world behavior occurs for $0<m<2$ while
in Ref. \cite{s7}, it was argued that small world behavior occurs
only for $m<1$. In the section $1<m<2$, according to Ref.
\cite{s7}, the average shortest-path length scales as $N^{\delta}$
with the value of the exponent $\delta$ $(0<\delta<\frac{1}{2})$
depending on $m$. Ref. \cite{s12} studied the critical behavior of
Ising model on such a one-dimensional network, since if the
network's behavior is that of small world it should be reflected
in the critical exponents of the Ising model, which will assume
mean-field values. Their result indicates that there is a
mean-field behavior for $m<1$ and a finite-dimensional behavior
for $1<m<2$, which is in accordance with the conclusion in Ref.
\cite{s7}.

In this paper, we re-examine the critical behavior of Ising model
on a one-dimensional network with distance-dependent connections
in Ref. \cite{s12} by using Monte Carlo simulations. But we
enlarge the underlying network sizes, and also take into account
the finite-size scaling analysis of the order parameter $[\langle
M\rangle]$ by using the plot of $[\langle M\rangle]N^{1/4}$ versus
$T$, which can reflect the mean-field nature indirectly \cite{s8}.
Here, $[\cdots]$ and $\langle\cdots\rangle$ denote the thermal
average taken over Monte Carlo steps for equilibrium at each
temperature, and over different network realizations,
respectively. Thus more reliable results can be produced from our
simulations, since statistical fluctuations are greatly suppressed
in these large-scale networks, and the finite-size effects are
much more discernible in the Binder's cumulant $U_{N}$
\cite{s16,s17} and the specific heat $C_{v}$ than in the order
parameter $[\langle M\rangle]$ \cite{s8} . The results show that
the model has a mean-field nature in the whole range of $0<m<2$.

In the next section, we will introduce the model we use and the
Monte Carlo simulation. In section
\uppercase\expandafter{\romannumeral3}, the results and analysis
are given. Section \uppercase\expandafter{\romannumeral4} is the
conclusion.

\section{Ising model and Monte Carlo simulations}
The underlying network of the one-dimensional Ising model used in
this paper is based on the generalized WS model \cite{s4}. The
addition of connections considers the distance between vertices.
That means the vertex $j$ to which a connection is attached is not
randomly selected, but according to a distribution depending on
the distance $l$ from $i$ to $j$: $\Theta(l)\sim l^{-m}$. We start
with a one-dimensional regular lattice with $N=2K+1$ vertices and
$2k=2$ nearest neighbors. From $i=0$, we select the $i$-th vertex
and generate a random number $\varphi\in[0,1)$ from a uniform
distribution. If $\varphi<p$, we select one vertex from the
clockwise $K$ vertices to attach to according to the probability
distribution $\Theta(l)\backsim l^{-m}$. Self-connections and
multiple connections are prohibited. We repeat this process until
all the vertices are selected, which results in a network with
$N(1+p)$ connections on average.

\begin{figure}[th]
\centering \rotatebox{-90}{\includegraphics[width=3.5cm]{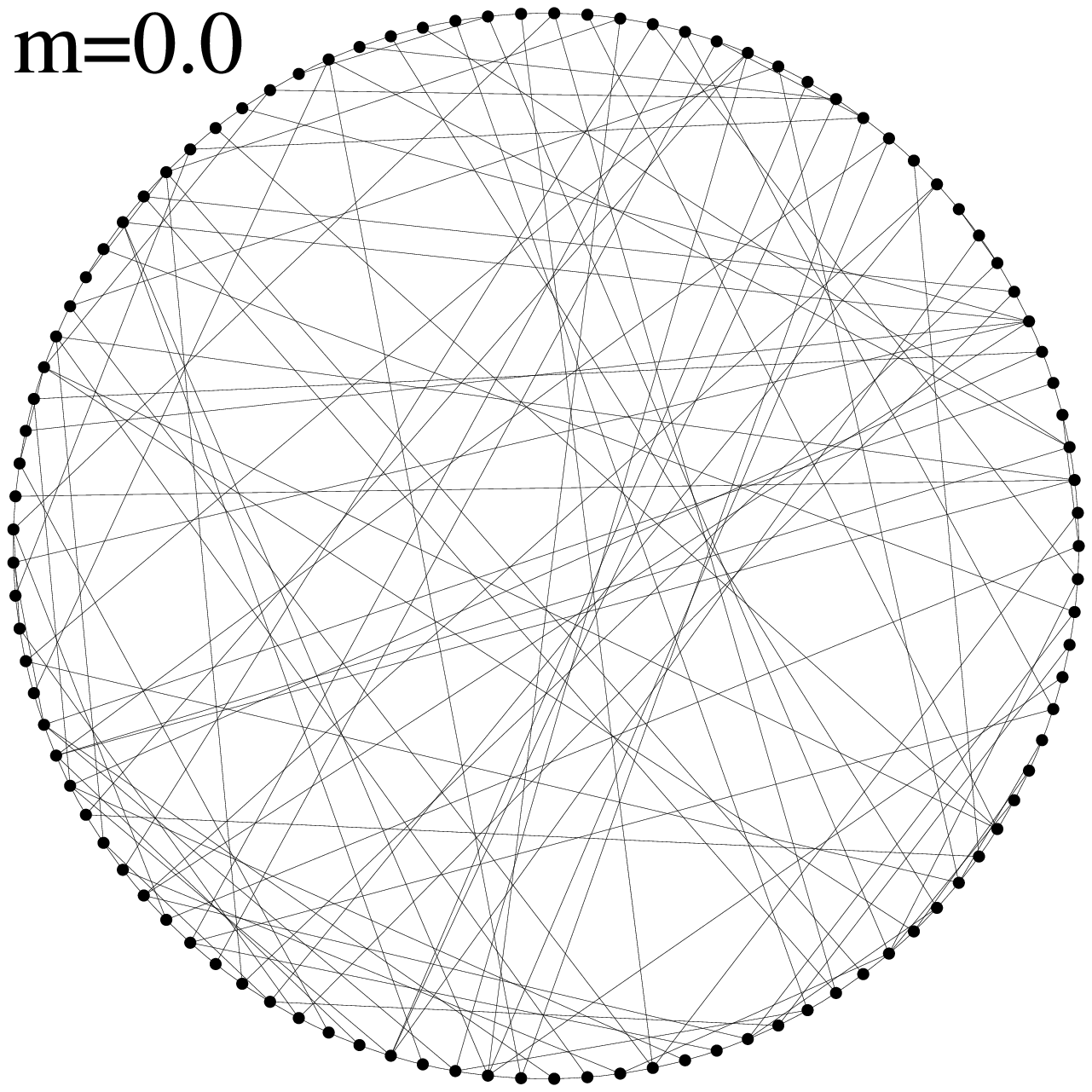}}
\vspace*{0pt}
\rotatebox{-90}{\includegraphics[width=3.5cm]{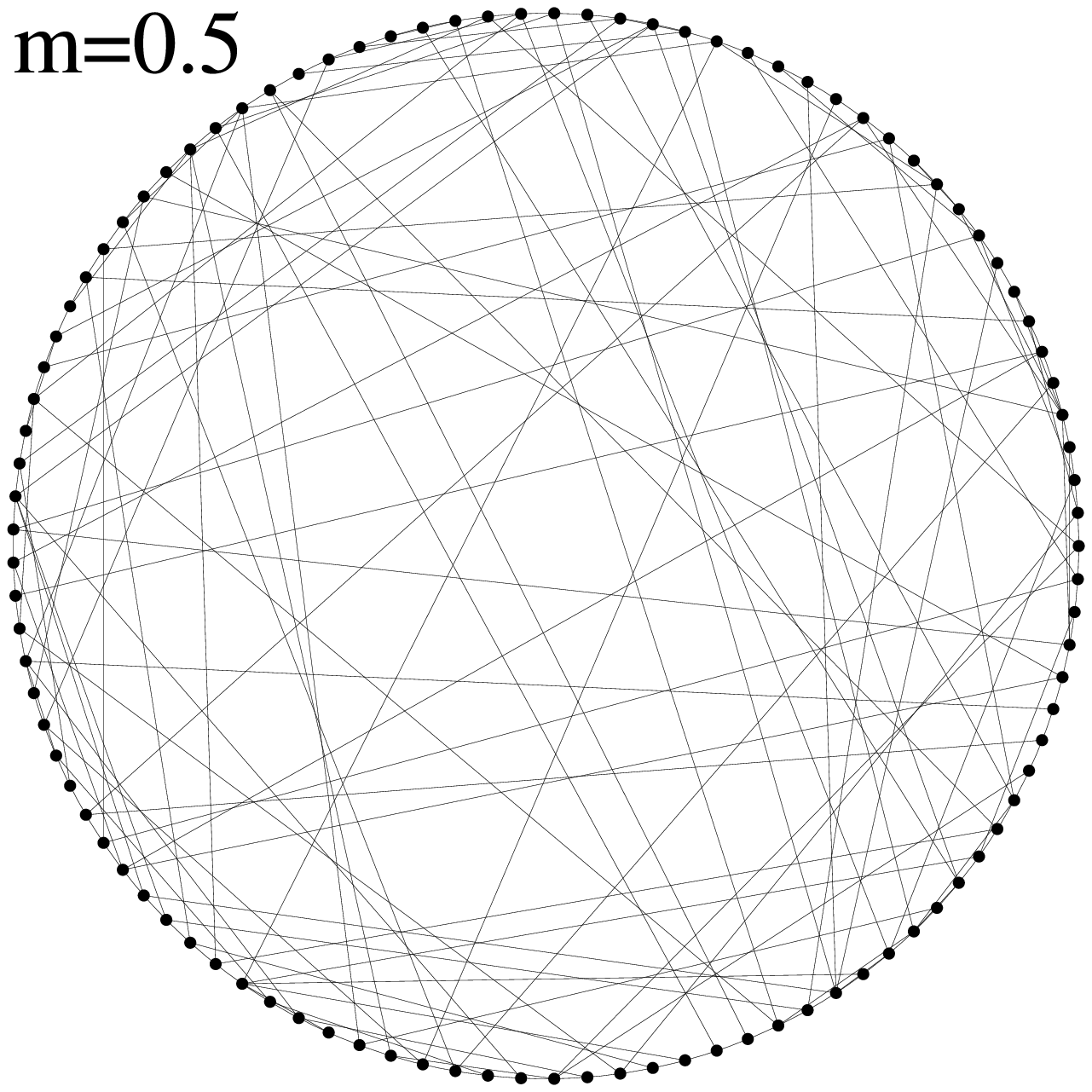}}
\vspace*{0pt}
\rotatebox{-90}{\includegraphics[width=3.5cm]{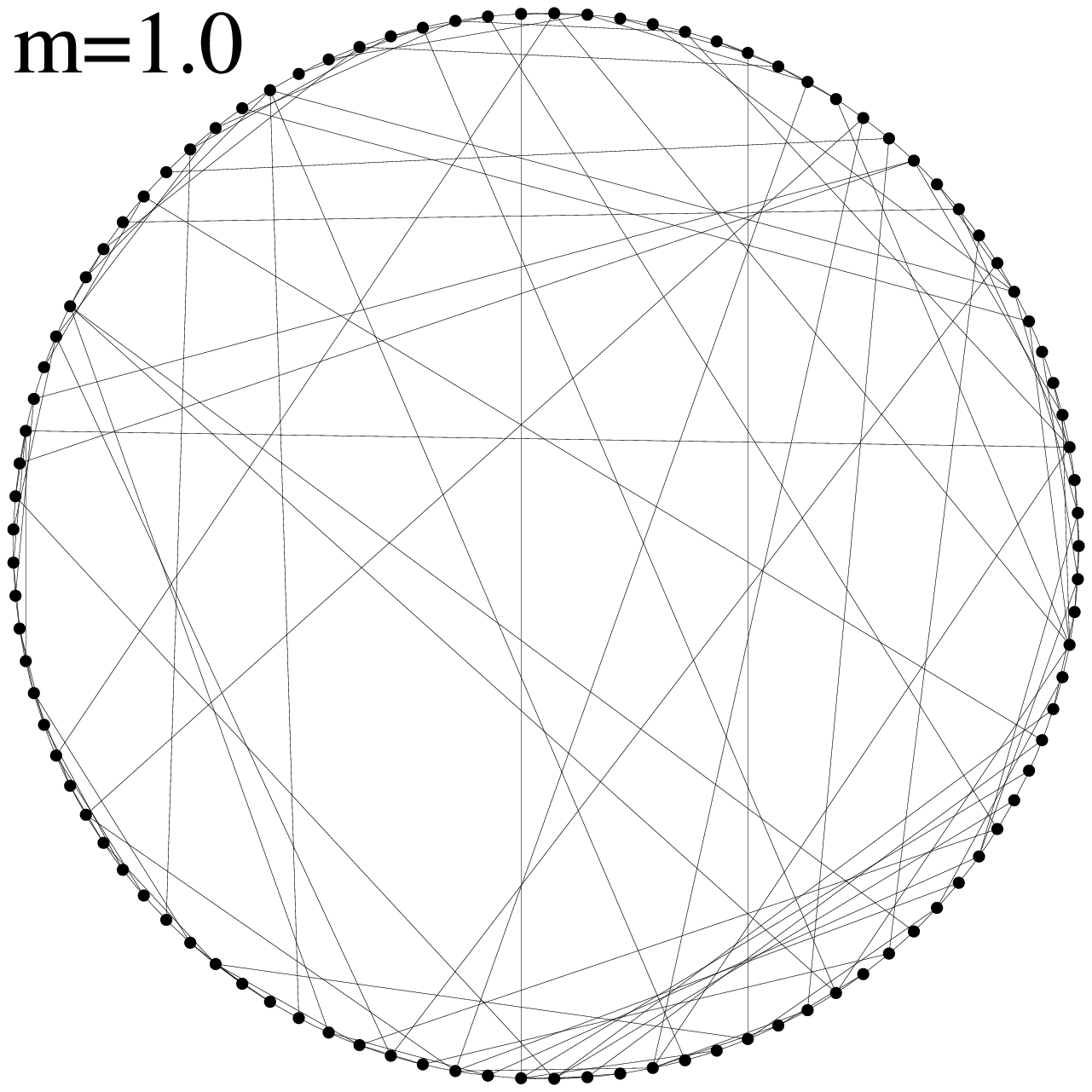}}
\vspace*{0pt}
\rotatebox{-90}{\includegraphics[width=3.5cm]{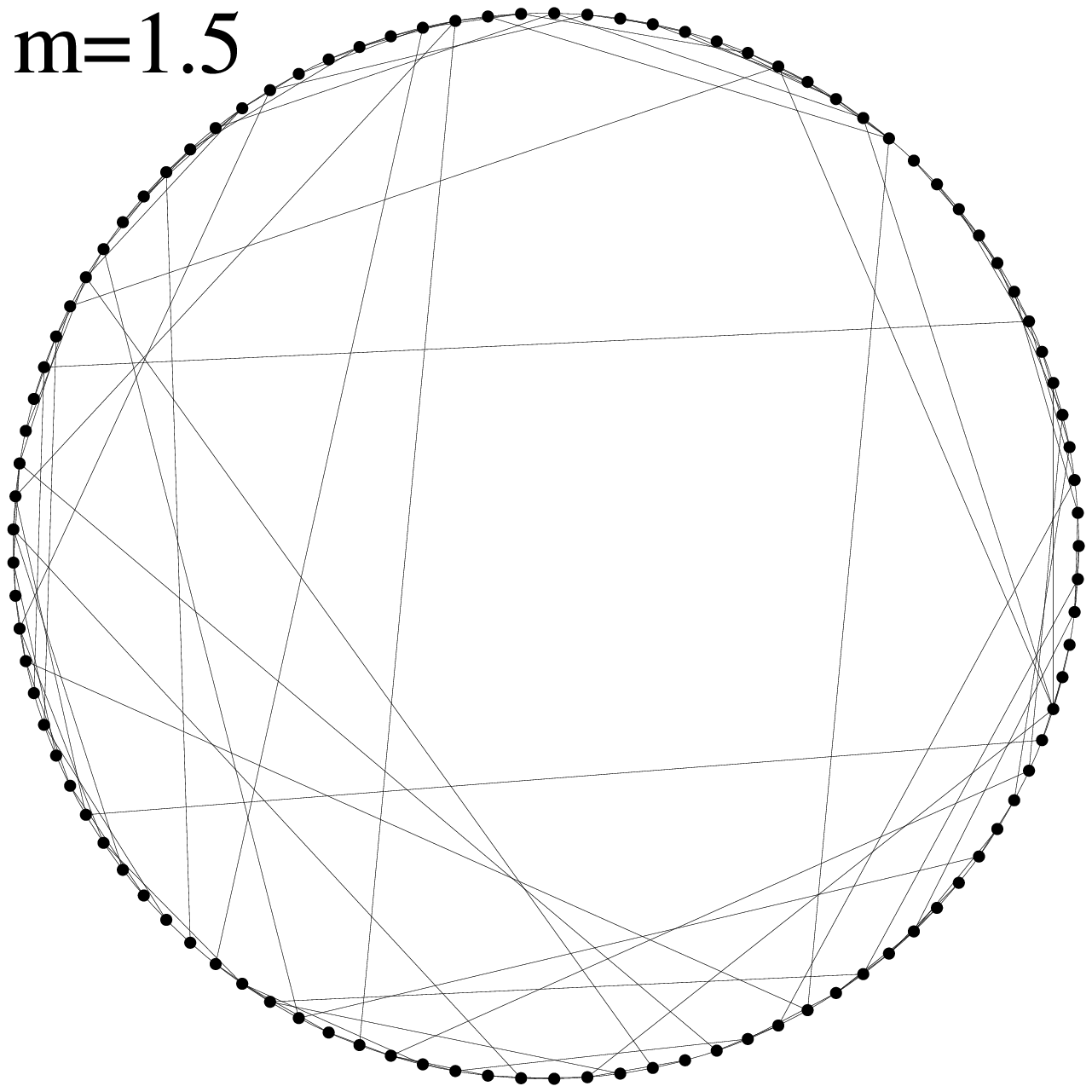}}
\vspace*{0pt}
\rotatebox{-90}{\includegraphics[width=3.5cm]{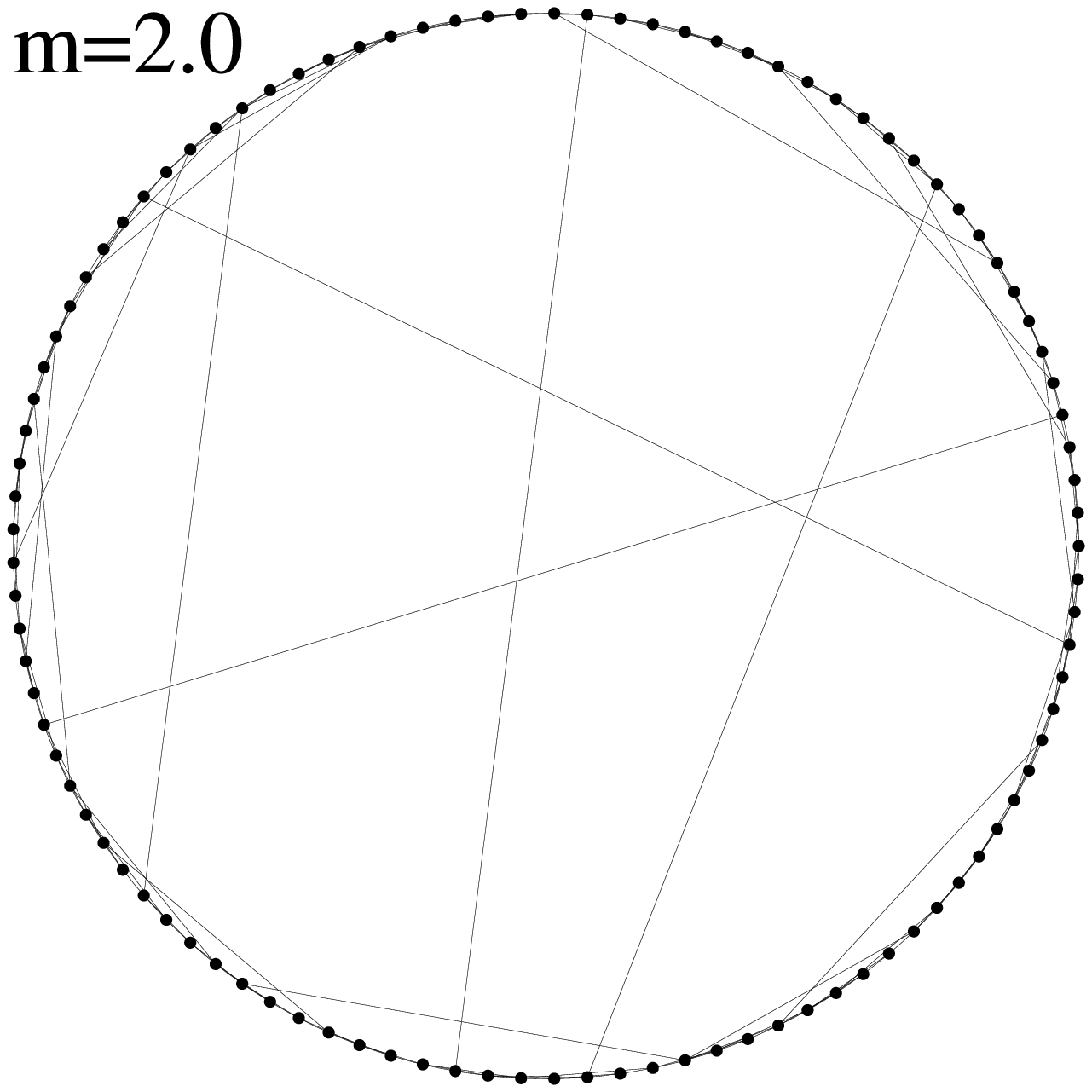}}
\vspace*{0pt}
\rotatebox{-90}{\includegraphics[width=3.5cm]{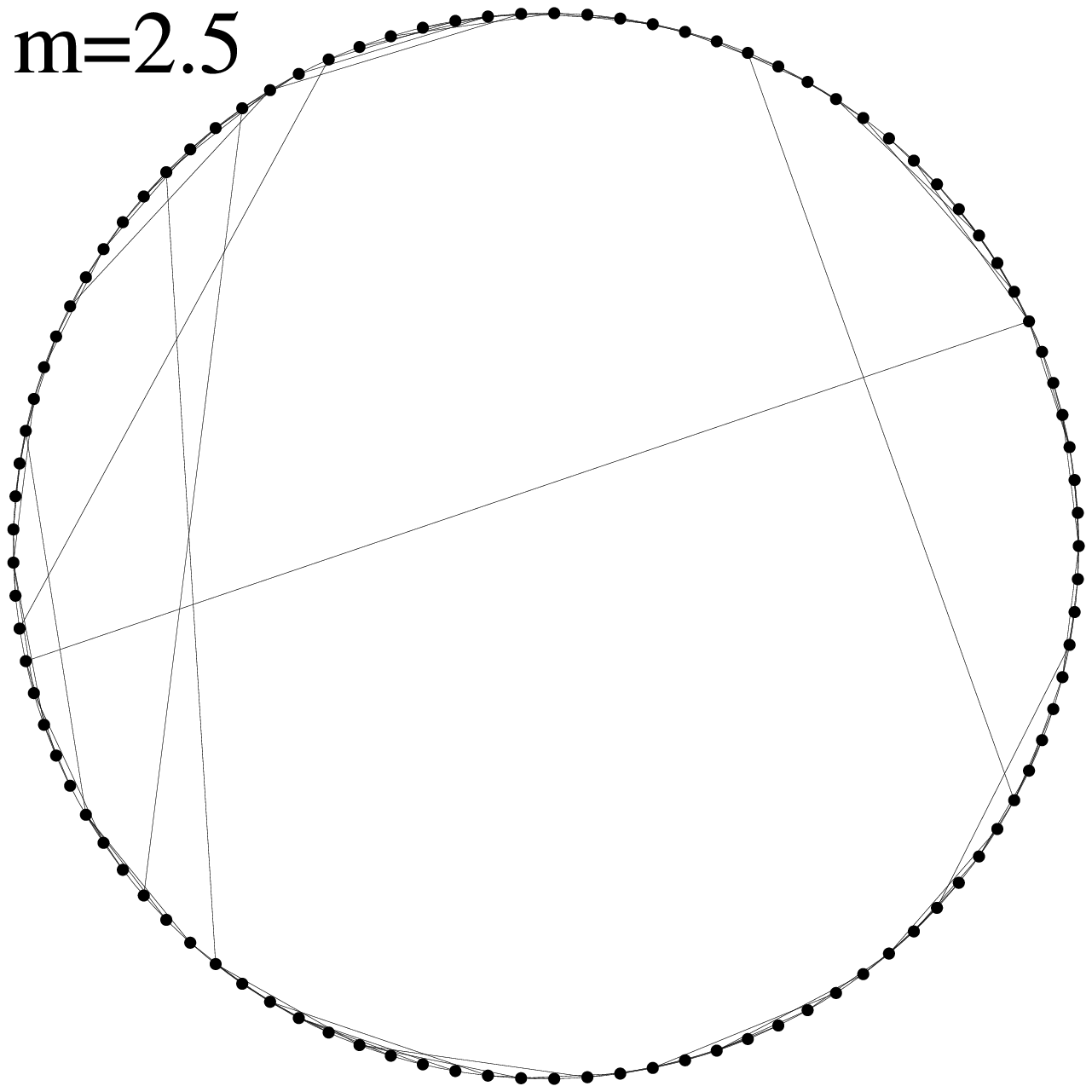}}
\vspace*{0pt} \caption{Examples of generated networks with $N=101$
under the same $p=1$ and different $m$ values.}
\end{figure}

In order to investigate the influence of $m$ on the critical
behavior of Ising model, in this paper we only consider the case
$p=1$ for simplicity. Our model with $m=0$ is equivalent to the
model used by Andrzej Pekalski \cite{s9}. As $m\rightarrow
\infty$, our model restores to the one-dimensional regular network
with $2k=4$ nearest neighbors. Fig.1 shows that with the increase
of $m$, long-range connections are more and more difficult to form
between vertices with long distances.

\begin{figure}[th]
\centering
\includegraphics[width=6cm]{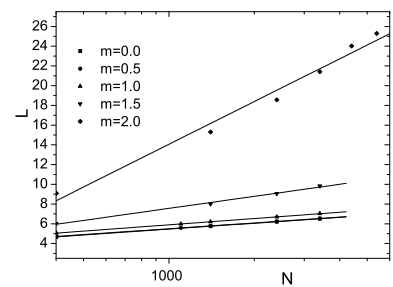}\vspace*{0pt}
\caption{The relationship of the average shortest-path length $L$
with the size of the network $N$ under different $m$ values (the
case $m=0.5$ is almost identical with $m=0$). All the plots have
been averaged over 50 realizations.}
\end{figure}

Fig.2 shows that the increase of $m$ results in the increase of
the average shortest-path length $L$, but the small world property
retains within the entire range $0<m<2$ as $L\varpropto\ln N$.
Fig.3(a) shows that with the increase of $m$ value, especially
after $m>1$, the network clustering coefficient becomes
independent of system size, which is the characteristic of regular
network. Furthermore, it can be seen from Fig.3(b) that, for a
certain network size, with the increase of $m$, the clustering
coefficient converges to $C_{max}=1/2$, which is identical to the
clustering coefficient of one-dimensional regular network with
$2k=4$ nearest neighbors.

\begin{figure}[th]
\centering
\includegraphics[width=4.8cm]{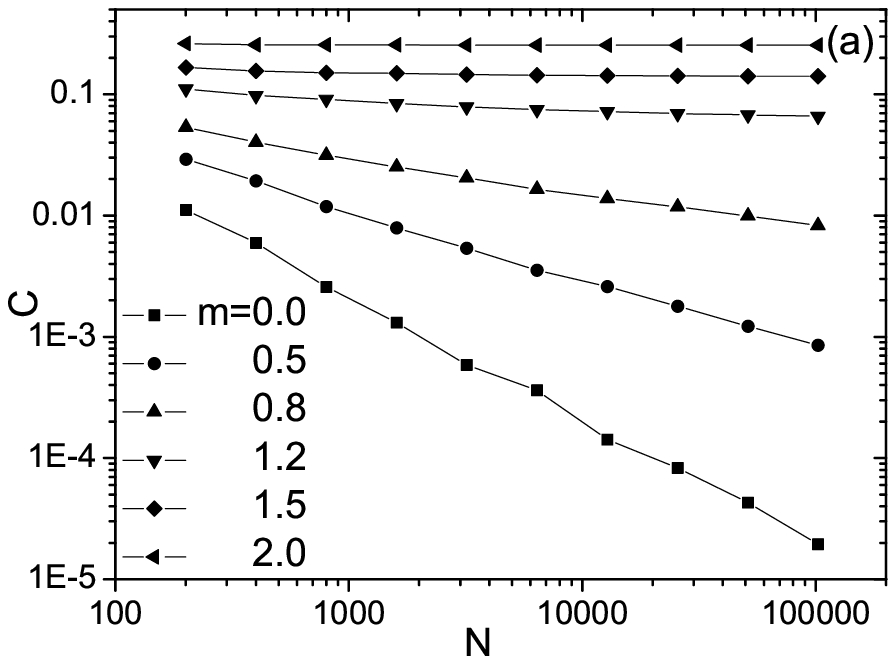}\hspace*{-5.0pt}
\hspace*{-15.0pt}\includegraphics[width=4.7cm]{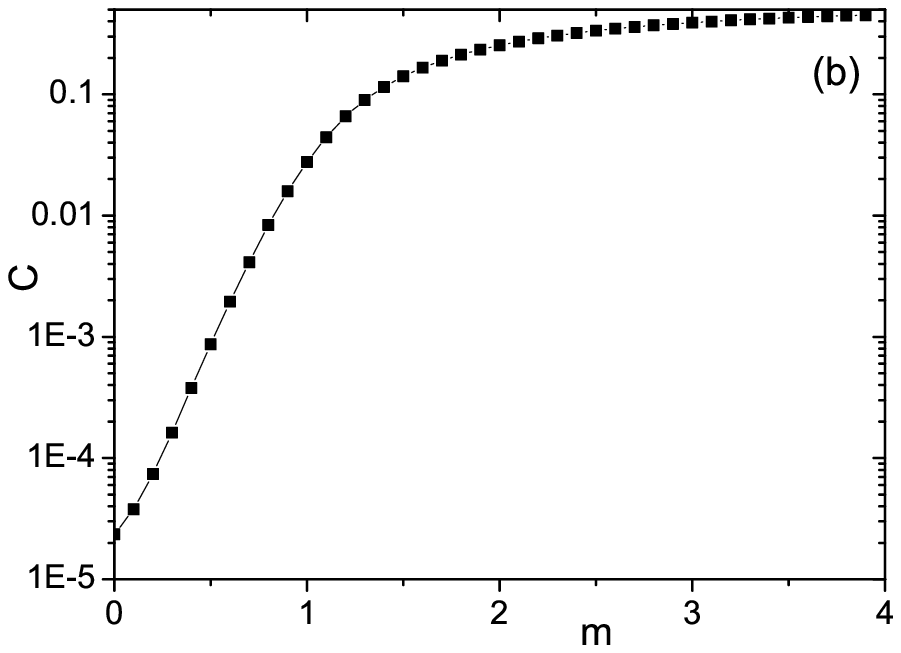}
\caption{(a) Network clustering coefficient $C$ versus network
size $N$ for different $m$ values. (b) Network clustering
coefficient $C$ versus $m$ value for a network with $N=100001$
vertices. The shown $C$ values in these two plots have been
averaged for 50 different network realizations.}
\end{figure}

The Ising model is described by the Hamiltonian:
\begin{equation}
H = -J \sum\limits_{(i,j)}  {\sigma}_i {\sigma}_j,
\end{equation}
where $J>0$ is the coupling constant between vertex $i$ and $j$ if
they are connected, $(i,j)$ is the collection of all the
connections in the network.

Our Monte Carlo simulation of the Ising model starts on a periodic
one-dimensional ring. At the beginning, all the spins on the ring
are given the same value $+1$, then a random vertex is selected to
flip according to the rule of the Metropolis algorithm
\cite{s9,s18} and Glauber dynamics \cite{s11} to simulate the
evolution of Ising model under different temperature. And
finite-size scaling is used to investigate the
paramagnetic-ferromagnetic transition temperature and the critical
exponents. During the simulations, we mainly use the random number
generator in \cite{s19}, and we also use the drand48() function in
standard C library as a comparison. It is found that there is no
notable influence on the simulation results. In the simulations we
measure the Binder'cumulant $U_{N}$, the susceptibility $\chi$ and
the specific heat $C_{v}$ \cite{s16,s17,s20,s21,s22}:
\begin{equation}
\begin{array}{lll}
\vspace*{0.2cm}
U_{N}=1-\dfrac{[\langle M^{4}\rangle]}{3[\langle M^{2}\rangle]^{2}},\\
\vspace*{0.2cm}
\chi=\dfrac{1}{N}\sum\limits_{ij}[\langle\sigma_{i}\sigma_{j}\rangle],\\
\vspace*{0.2cm} C_{v}=\dfrac{[\langle H^{2}\rangle-\langle
H\rangle^{2}]}{T^{2}N}
\end{array}
\end{equation}
with $M=|\frac{1}{N}\sum\limits_{i}\sigma_{i}|$. Here $[\cdots]$
denotes the thermal average taken over $2.5\times 10^{4}$ Monte
Carlo steps after discarding the initial $2.5\times 10^{4}$ ones
for equilibrium at each temperature, and $\langle\cdots\rangle$
denotes the average over different network realizations taken over
$30-80$ configurations.

Besides these three physical quantities, the finite-size scaling
analysis of the order parameter $[\langle M\rangle]$, which
exhibits the critical behavior $[\langle
M\rangle]\sim(T_{c}-T)^{\beta}$ in the thermodynamic limit, is
used. In a finite-sized system, the order parameter scales as
\cite{s8}:
\begin{equation}
[\langle
M\rangle]=N^{-\beta/\bar{\nu}}g((T-T_{c})N^{1/\bar{\nu}}),
\end{equation}
where $g(x)$ is an appropriate scaling function. Eq.(4) leads to a
unique crossing point at $T_{c}$ in the plot of $[\langle
M\rangle]N^{\beta/\bar{\nu}}$ versus $T$, so Beom et al \cite{s8}
suggested analyzing the finite-size scaling of the order parameter
by $[\langle M\rangle]N^{1/4}$ for large $m$ to overcome the
finite-size effects, which are more prominent in other
thermodynamic quantities.

\section{Results and Analysis}

\begin{figure}[th]
\centering
\includegraphics[width=4.6cm]{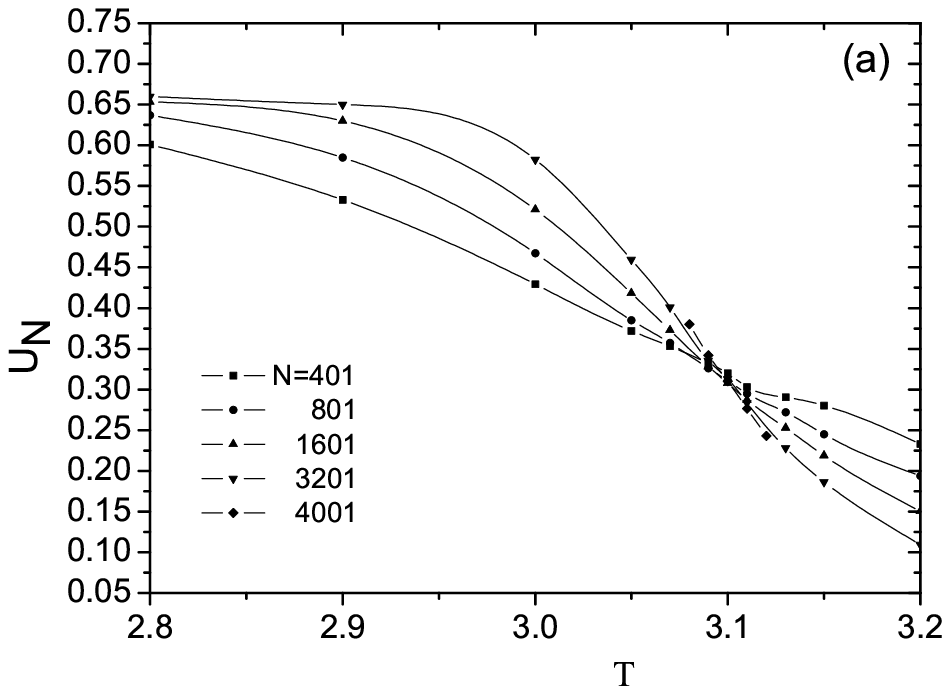}\vspace*{-8.0pt}
\hspace*{-20.0pt}\includegraphics[width=4.6cm]{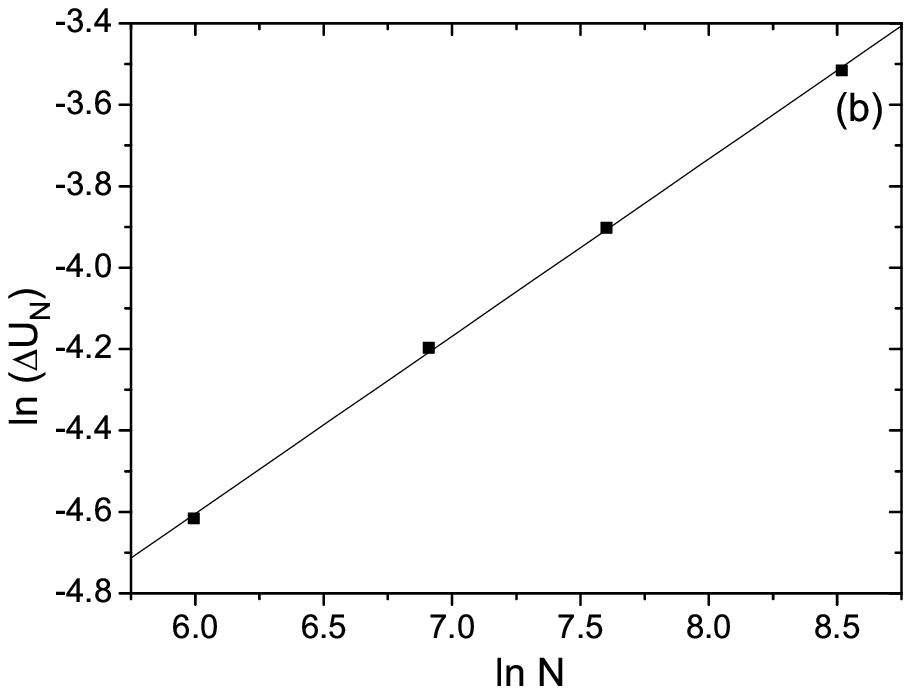}\vspace*{-8.0pt}
\includegraphics[width=4.6cm]{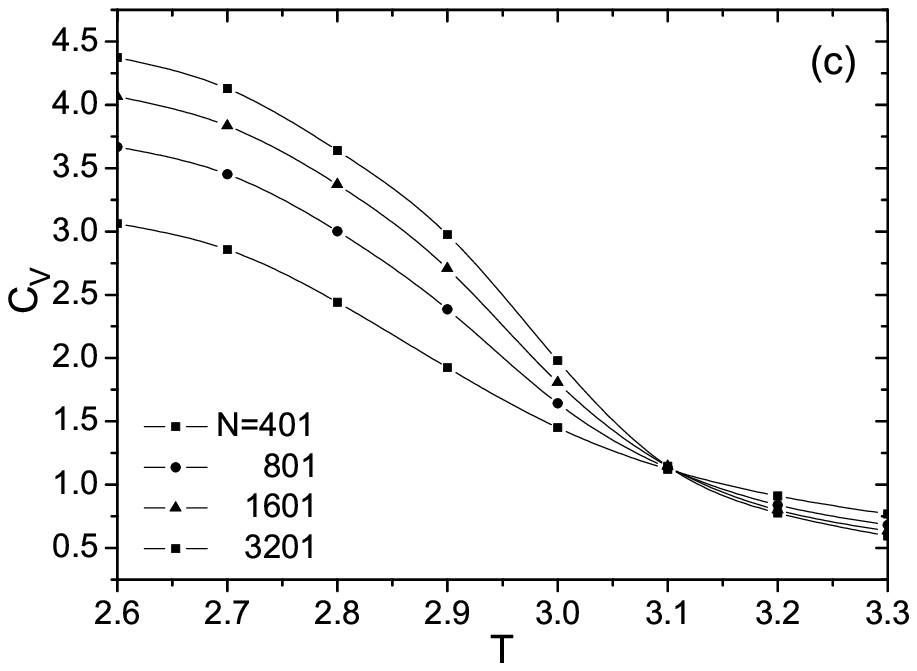}\vspace*{-8.0pt}
\hspace*{-20.0pt}\includegraphics[width=4.6cm]{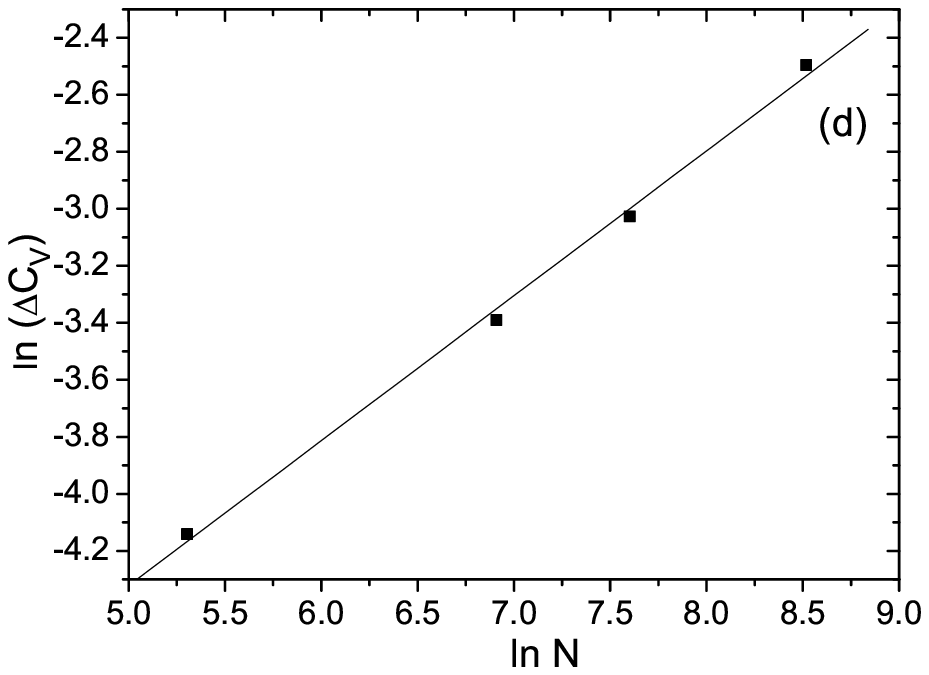}\vspace*{-8.0pt}
\includegraphics[width=4.6cm]{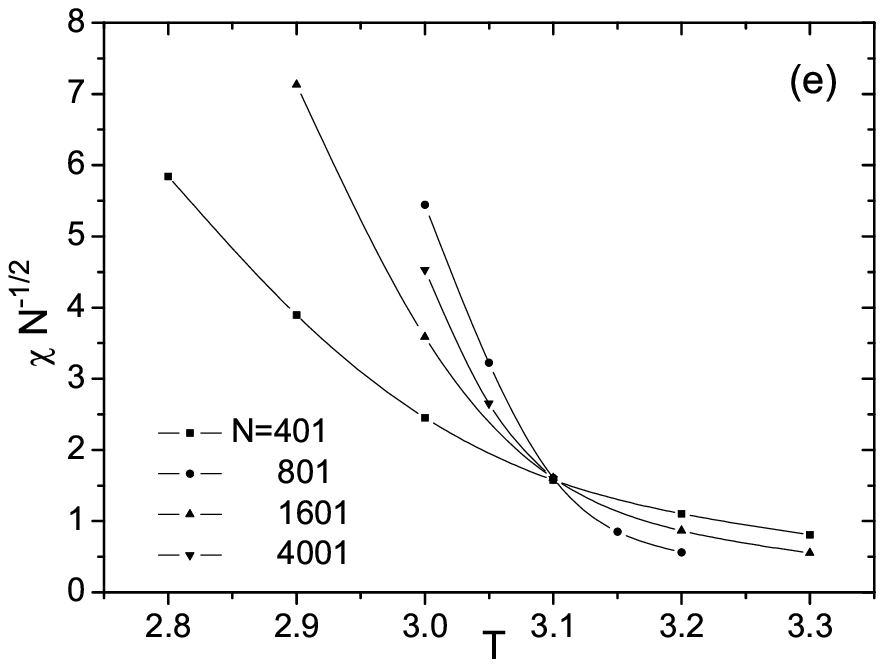}\vspace*{-8.0pt}
\hspace*{-20.0pt}\includegraphics[width=4.6cm]{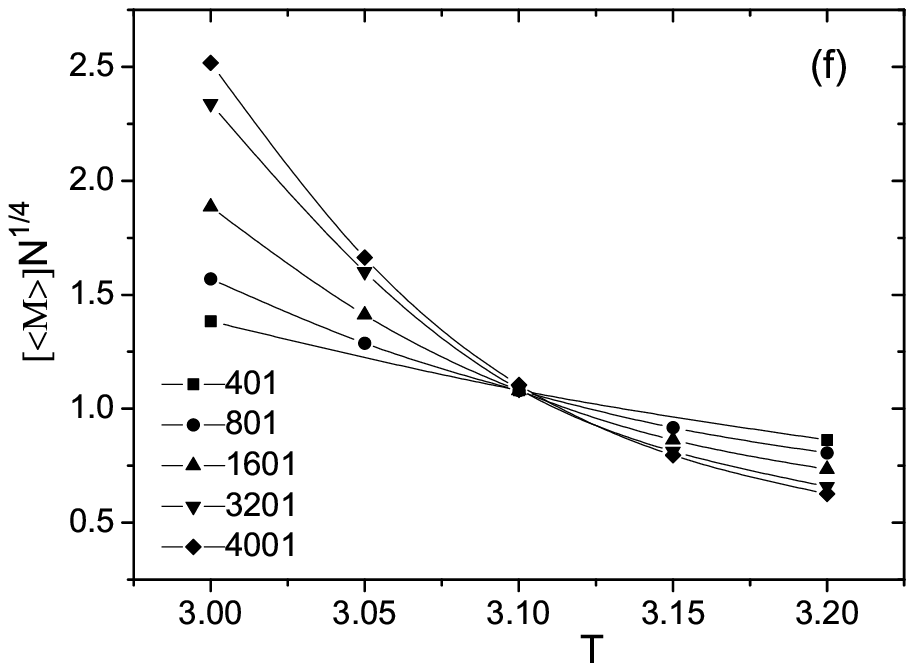}
\caption{For $m=0$, (a) Binder's cumulant $U_{N}$ has a unique
crossing point at $T_{c}=3.10(5)$(in units of $J/k_{B})$. (b) The
critical exponent $\overline{\nu}=2.08$ is obtained from the
linear fit of $\ln\Delta U_{N}$ vs. $\ln N$. (c) Specific heat
$C_{v}$ has a unique crossing point at $T_{c}=3.10(7)$, which
indicates $\alpha=0$. (d) The expansion of $C_{v}$ near $T_{c}$
obtains $\overline{\nu}=2.04$. (e) Finite-size scaling of the
susceptibility again determines $T_{c}=3.10(5)$ with the critical
exponent $\gamma=1$. (f) Finite-size scaling of the order
parameter $[\langle M\rangle]$ leads to a unique crossing point at
$T_{c}=3.10(1)$, and yields $\beta=\frac{1}{2}$, $\bar{\nu}=2.0$
(see Eq.(4) \cite{s8}).}
\end{figure}

As shown in Fig.4, for $m=0$, the finite-size scaling of the
Binder's cumulant $U_{N}$, the specific heat $C_{v}$, the
susceptibility $\chi$ and the order parameter $[\langle M\rangle]$
all reveal unanimously $T_{c}=3.10(5)$ (in units of $J/k_{B}$),
which confirms the presence of a finite-temperature transition.
And the obtained critical exponents indicate the mean-field nature
of the transition.

Till now, the most controversial property of the phase transition
of one-dimensional Ising model with distance-dependent connections
given by Eq.(1) is in the range of $1\leq m<2$ \cite{s6,s7,s12},
so we will focus on the critical behavior of Ising model for
$1\leq m<2$.

\begin{figure}[th]
\centering
\includegraphics[width=4.6cm]{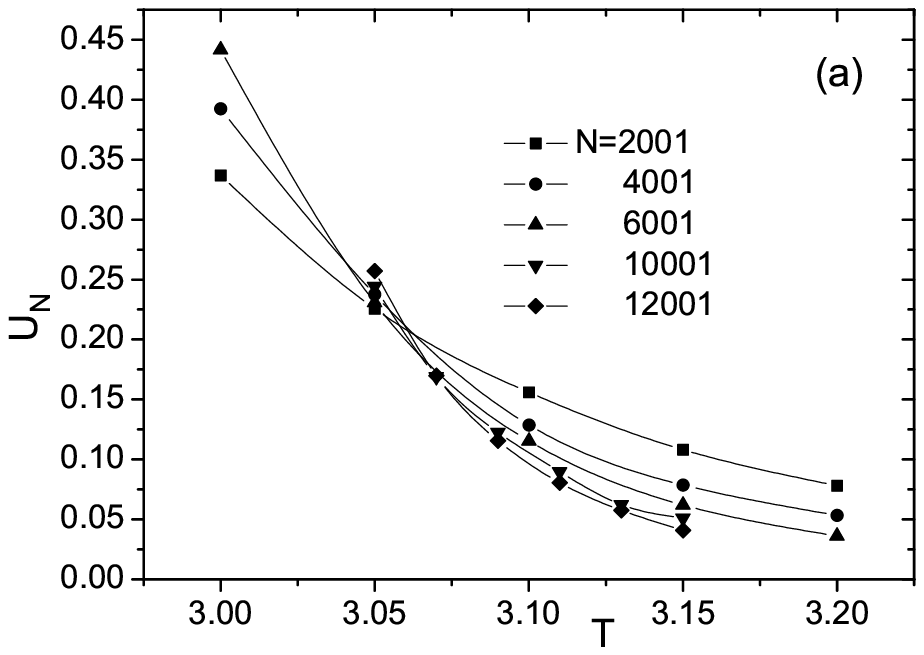}\vspace*{-8.0pt}
\hspace*{-20.0pt}\includegraphics[width=4.6cm]{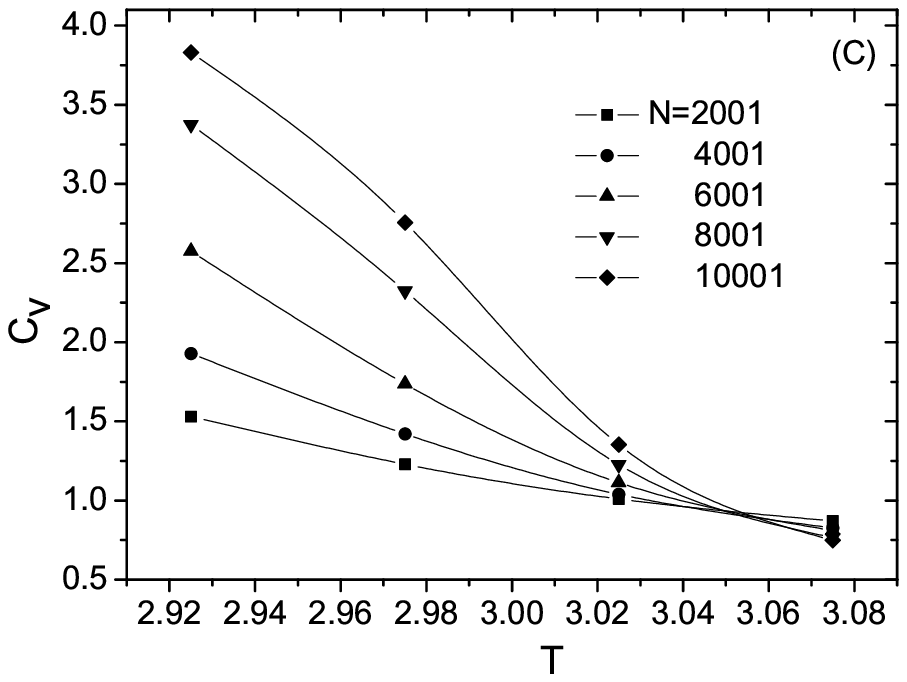}\vspace*{-8.0pt}
\includegraphics[width=4.6cm]{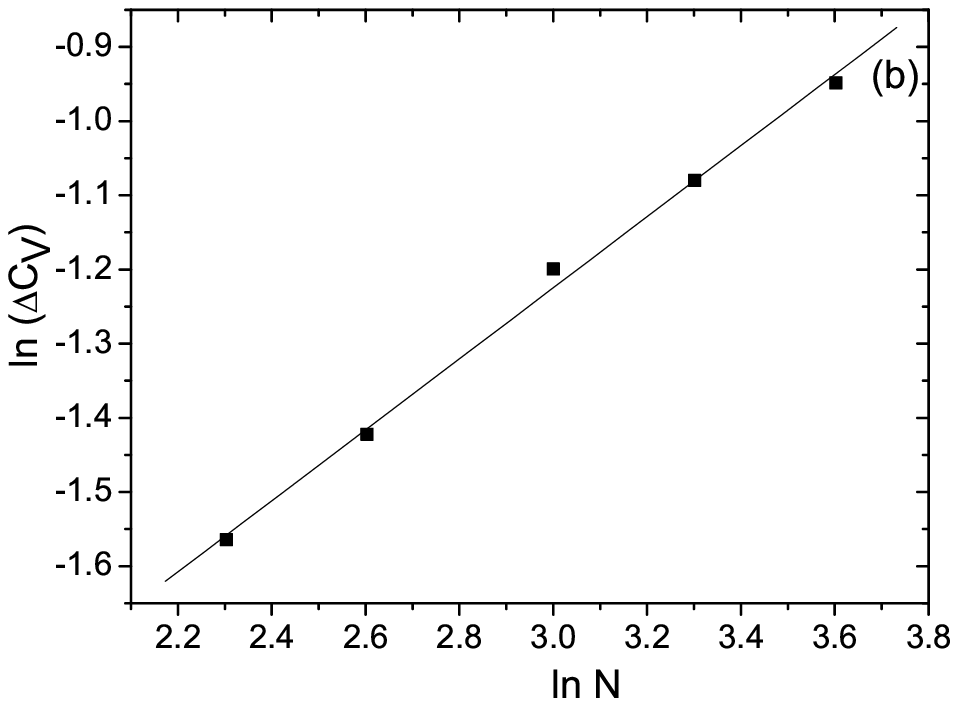}\vspace*{-8.0pt}
\hspace*{-20.0pt}\includegraphics[width=4.6cm]{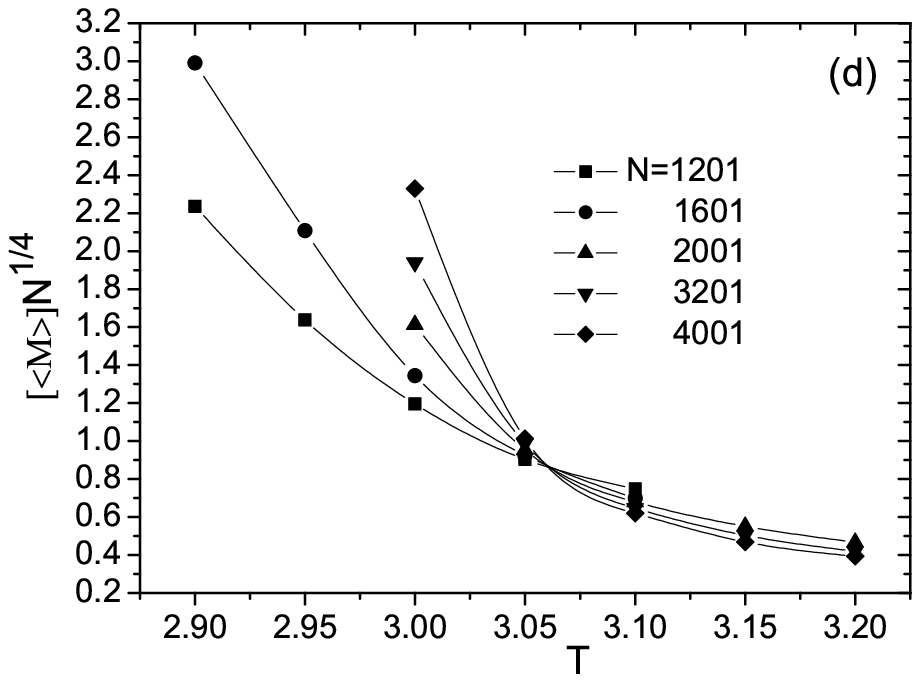}
\caption{For $m=1.0$, (a) Finite-size scaling of the Binder's
cumulant $U_{N}$ has a unique crossing point at $T_{c}=3.05(2)$.
(b) Finite-size scaling of the specific heat $C_{v}$ determines
$T_{c}=3.05(3)$, which indicates $\alpha=0$. (c) The expansion of
$C_{v}$ near $T_{c}$ obtains $\overline{\nu}=2.05$. (d) Finite
size-scaing of the order parameter $[\langle M\rangle]$ leads to a
unique crossing point at $T_{c}=3.05(3)$, and yields
$\beta=\frac{1}{2}, \bar{\nu}=2$ (see Eq.(4) \cite{s8}).}
\end{figure}

Fig.5 presents the finite-size scaling of the Binder's cumulant
$U_{N}$, the specific heat $C_{v}$ and the order parameter
$[\langle M\rangle]$ at $m=1.0$. The measured quantities remain a
unique crossing point and reveal unanimously $T_{c}=3.05(3)$,
which confirms the presence of a finite-temperature phase
transition. The obtained critical exponents assume the mean-field
values as well.

\begin{figure}[th]
\centering
\includegraphics[width=4.6cm]{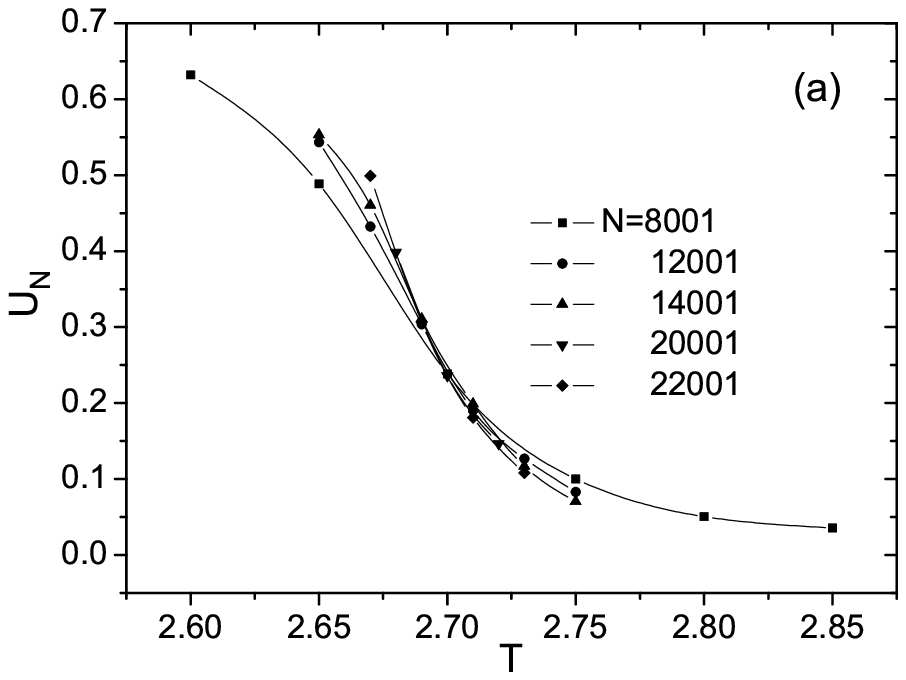}\vspace*{-8.0pt}
\hspace*{-20.0pt}\includegraphics[width=4.6cm]{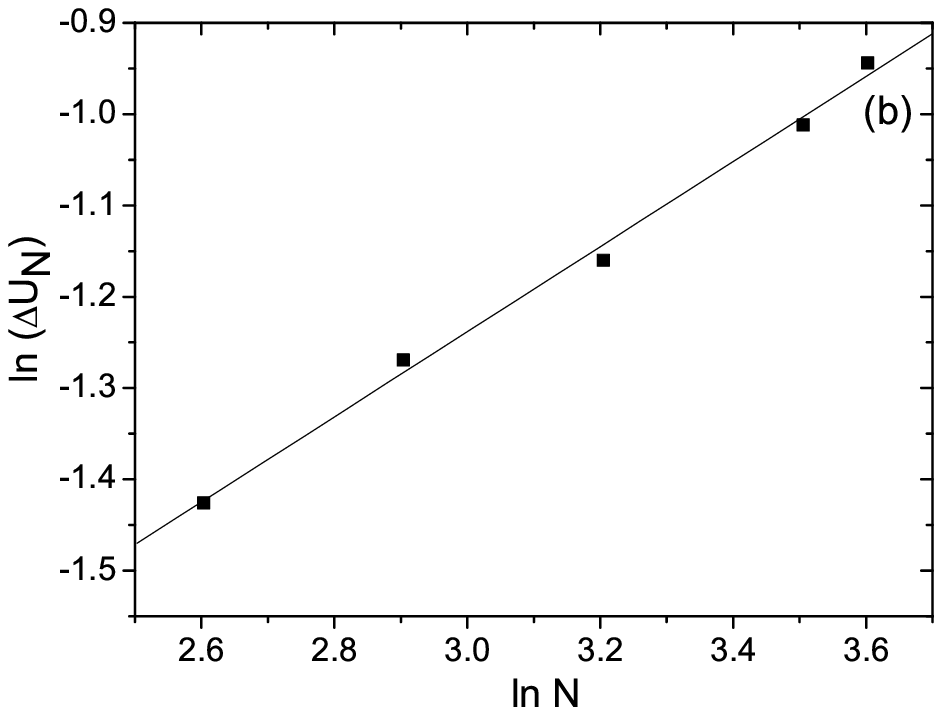}\vspace*{-8.0pt}
\includegraphics[width=4.6cm]{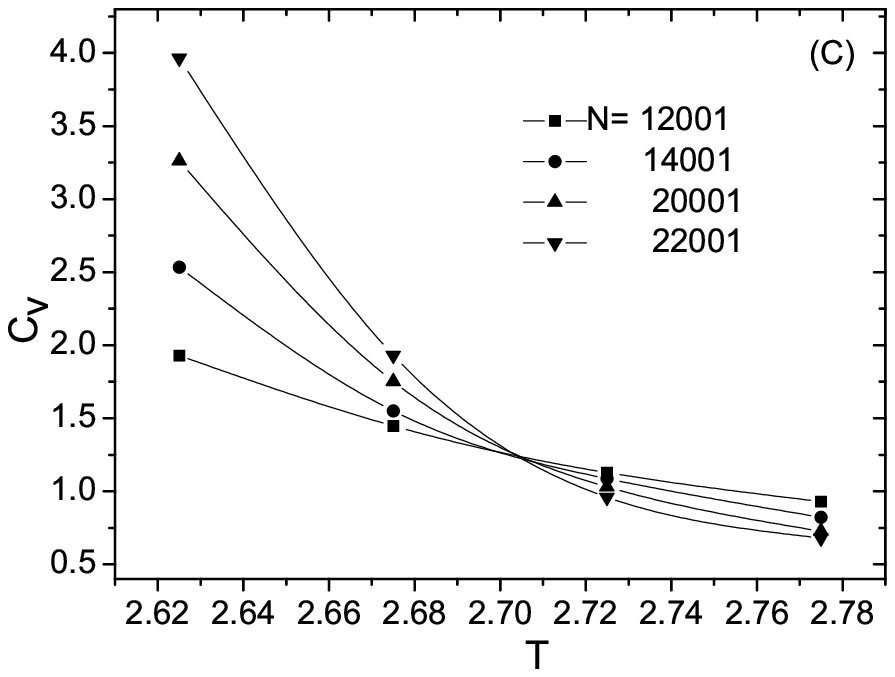}\vspace*{-8.0pt}
\hspace*{-20.0pt}\includegraphics[width=4.6cm]{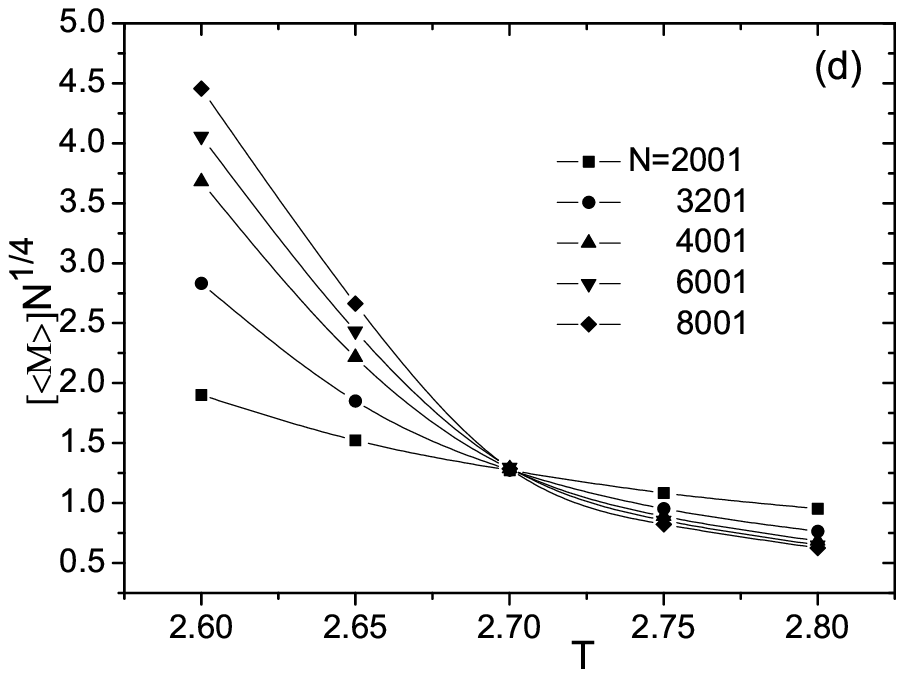}
\caption{For $m=1.5$, (a) Finite-size scaling of the Binder's
cumulant $U_{N}$ has a unique crossing point at $T_{c}=2.70(2)$.
(b) The expansion of $U_{N}$ near $T_{c}$ obtains
$\overline{\nu}=2.03$. (c)Finite-size scaling of the specific heat
$C_{v}$ determines $T_{c}=2.70(4)$, which indicates $\alpha=0$.
(d) Finite size-scaing of the order parameter $[\langle M\rangle]$
leads to a unique crossing point at $T_{c}=2.70(2)$, and yields
$\beta=\frac{1}{2}, \bar{\nu}=2$ (see Eq.(4) \cite{s8}).}
\end{figure}

Fig.6 is the finite-size scaling of the Binder's cumulant $U_{N}$,
the specific heat $C_{v}$ and the order parameter $[\langle
M\rangle]$ at $m=1.5$. The curves of Binder's cumulant $U_{N}$ do
not intersect to one unique crossing point very well. But the
trends of the evolution indicate that it will emerge when the
underlying network size is large enough. the finite-size scaling
of the order parameter $[\langle M\rangle]$ under different
network sizes behaves better in intersecting to one unique point,
revealing unanimously $T_{c}=2.70(2)$. The obtained critical
exponents values indicate the presence of a mean-field
finite-temperature phase transition.

\begin{figure}[th]
\centering
\includegraphics[width=4.6cm]{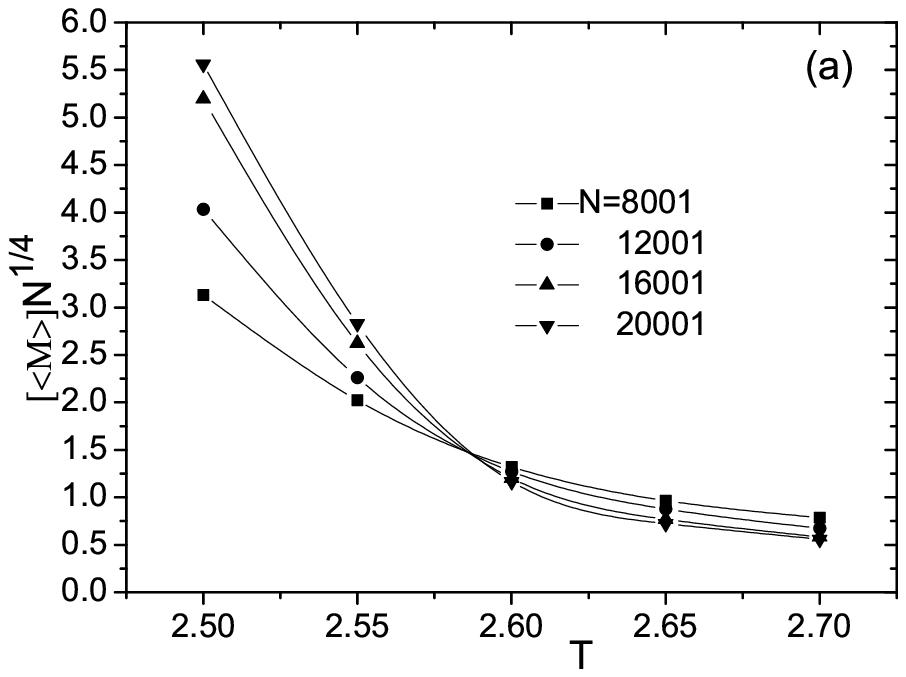}\vspace*{-8.0pt}
\hspace*{-20.0pt}\includegraphics[width=4.6cm]{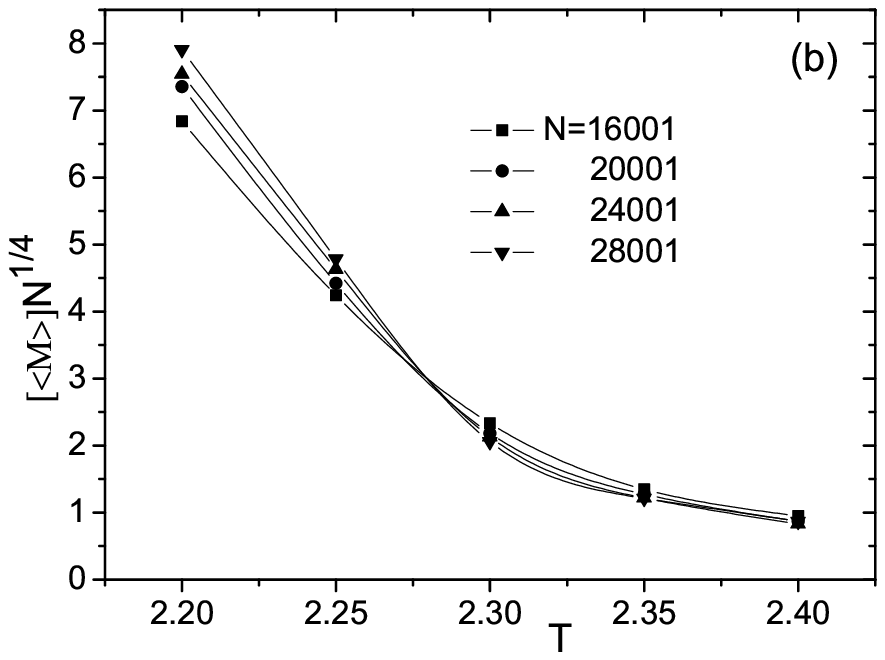}
\caption{Finite-size scaling of the order parameter $[\langle
M\rangle]$ gives $T_{c}=2.58(5)$ at $m=1.6$ (a)  and
$T_{c}=2.28(2)$ at $m=1.8$ (b).}
\end{figure}

\begin{figure}[th]
\centering
\includegraphics[width=6cm]{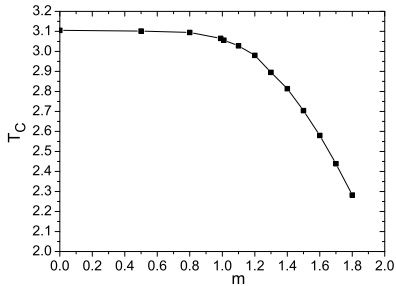} \vspace*{0pt}
\caption{The evolution of the transition temperature with the
increase of $m$.}
\end{figure}

With the increase of $m$, the algebra decreasing of $\Theta(l)\sim
l^{-m}$ will will have a cut-off due to the finite network size.
This makes the normalization of the distribution function depend
on the size of the network, and makes it inevitable to study
systems of very large sizes so as to obtain correct scaling
behavior \cite{s8,s22}. By analyzing the plot of $[\langle
M\rangle]N^{1/4}$ versus $T$, we get further progress on the
transition temperature of systems with large scale, as shown in
Fig.7 as examples. This also indicates the fact that the phase
transition still holds a mean-field nature for $m>1$. The phase
transition temperatures here are higher than those found by H.
Hong et al in Ref. \cite{s20}, which may be due to the presence of
more long-range connections in our model which contribute to
different phase transition properties.

In Fig.8, the decrease of $T_{c}$ in the range of $[0,1)$ is not
significant, but this decreasing becomes obvious after $m>1$. This
brings difficulties to the phase transition analysis of Ising
model due to the following two factors: a) With the increase of
$m$, the appearance of a convincing unique crossing point needs
larger network sizes, as shown in Fig.6 and Fig.7; b) The
transition temperature decreases rapidly after $m>1$ (Fig.8), so
we need to do Monte Carlo simulation at lower temperature, which
means lower probability to flip. What's more, the value of the
order parameter $[\langle M\rangle]$ has a strong correlation with
its value at the previous step. Thus the system needs more Monte
Carlo steps to reach equilibrium, and we need more iterations to
get the thermodynamic average. Thus the simulations will be more
time-consuming.

\section{Conclusion}
In this paper, we reconsider the phase transition of Ising model
on a one-dimensional network with distance-dependent connections
given by $\Theta(l)\backsim l^{-m}$, by adding the finite-size
scaling analysis of the order parameter $[\langle M\rangle]$ to
overcome the difficulties at high $m$ values. Ref.\cite{s12} based
their model on networks with hundreds of vertices and concluded
that the phase transition has a mean-field nature only in the
range of $0<m<1$. Our results, based on larger network sizes and
further finite-size scaling analysis of the order parameter
$[\langle M\rangle]$ as well as the Biner's cumulant $U_{N}$, the
specific heat $C_{v}$ and the susceptibility $\chi$, show that the
phase transition has a mean-field nature in the whole range of
$0<m<2$.

Because of the constraint of the long simulation time, we can not
apply the analysis to networks with even larger sizes at present.
Thus we can not make sure whether there is a critical value of $m$
at or beyond $m=2$, above which there will be no mean-field nature
or finite-temperature phase transition since the transition point
$m=2$ was suggested by the study of the network topological
structure \cite{s6,s7,s12}. With the finding of a proper solution
to the long simulation time, more interesting results are supposed
to come out.

\begin{acknowledgments}
We wish to acknowledge the financial support of the National
Natural Science Foundation of China under grant no. 70571027,
70401020, 10647125 and 10635020, and the Ministry of Education of
China under grant no. 306022.
\end{acknowledgments}

\end{document}